\documentstyle[12pt]{article}
\newcommand{\be}{\begin{equation}}
\newcommand{\ee}{\end{equation}}
\newcommand{\bea}{\begin{eqnarray}}
\newcommand{\eea}{\end{eqnarray}}
\newcommand{\bdm}{\begin{displaymath}}
\newcommand{\edm}{\end{displaymath}}

\newcommand{\cod}{d^{\dagger}}
\newcommand{\we}{\wedge}

\newcommand{\Mfold}{(M,g)}
\newcommand{\Hi}{{\cal H}_i}
\newcommand{\Sinf}{S^2_{\infty}}
\newcommand{\ginf}{g_{\infty}}
\newcommand{\finf}{f_{\infty}}
\newcommand{\gtens}{\mbox{\boldmath $g$}}
\newcommand{\htens}{\mbox{\boldmath $h$}}
\newcommand{\gtiltens}{\tilde{\gtens}}
\newcommand{\etatil}{\tilde{\eta}}
\newcommand{\doc}{\langle\!\langle {\cal J} \rangle\!\rangle}
\newcommand{\sprod}[2]{\langle #1 \, , #2 \rangle}

\begin{document}

\begin{center}
{\large\bf On the Uniqueness of the} \\
{\large\bf Papapetrou--Majumdar Metric} \\
\vspace{.5cm}
{\bf Markus Heusler}  \\
\vspace{.5cm}
Institute for Theoretical Physics \\
The University of Zurich \\
CH--8057 Zurich, Switzerland \\
\end{center}

\vspace{1.cm}

\begin{center}
{\bf Abstract}
\end{center}

\begin{quote}
We establish the equality of the ADM mass and the 
total electric charge for asymptotically flat, static 
electrovac black hole spacetimes with completely 
degenerate, not necessarily connected horizons.
\end{quote}

\vspace{1.cm}

\noindent
In a recent letter \cite{CQG12-L17-95}, Chru\'sciel 
and Nadirashvili have established the uniqueness of 
the standard Pa\-pa\-pe\-trou--Ma\-jum\-dar (PM)
spacetimes \cite{PRIAA51-191-45}, \cite{PR72-390-47} 
amongst the non--singular (in an appropriate sense, see 
\cite{CQG12-L17-95} for detailed conditions) electrovac 
black hole spacetimes of the PM form \cite{JMP13-865-72}, 
\cite{PRL27-1668-71}, \cite{CMP26-87-72}.
As the authors point out, a further step towards a 
complete classification of the static, asymptotically 
flat electrovac spacetimes could be achieved by proving 
the equality $M = |Q|$, which is supposed to hold if 
the horizon has (at least some) degenerate components. 
The aim of this letter is to derive this relation between 
the ADM mass and the total electric charge under the 
additional assumptions that all components of the 
Killing horizon have vanishing surface gravity and that 
the charges assigned to these components 
(see eq. (\ref{MiQi}) for the precise definition) 
have the same sign.

Let us consider a static, asymptotically flat electrovac 
spacetime $\Mfold$ with a not necessarily connected 
Killing horizon $H$. The metric assumes locally the form
\be
\gtens \, = \, - \, V \, dt \otimes dt \, + \, \gtiltens \; ,
\; \; \; \; \mbox{where} \; \; V \, \equiv \, - \, \sprod{k}{k}
\label{metric}
\ee
and where $\langle \cdot \, ,  \cdot \rangle$ denotes the
inner product with respect to $\gtens$. The Killing field 
($1$-form) $k$ is timelike ($V > 0$) throughout the domain 
of outer communications $\doc$, and the potential $V$ vanishes 
only on $H$, i.e., on the null hypersurface to which $k$ is 
orthogonal. 

We shall also use the fact that each connected 
component $\Hi$ of the intersection of $H$ with
an asymptotically flat, spacelike hypersurface $\Sigma$
is a topological $2$--sphere, provided that $\doc$ is 
simply connected. In fact, the 'topological censorship' 
hypothesis has meanwhile been established, mainly on the basis 
of global hyperbolicity and the null energy condition (see
\cite{CQG11-L147-94}, \cite{CMP151-53-93},
\cite{DGMP-G-94}, \cite{DGMP-C-94}, and
\cite{CQG-13-96} for new results and more references).

In a stationary spacetime the total mass $M$ and 
the charge $Q$ (as seen from the asymptotically flat 
region) are given by the flux integrals
\be
M \, = \, - \frac{1}{8 \pi} \int_{\Sinf} \ast dk \; ,
\; \; \; \;
Q \, = \, - \frac{1}{4 \pi} \int_{\Sinf} \ast F \; ,
\label{MQ}
\ee
where $F$ denotes the electromagnetic $2$--form and 
$\ast$ is the Hodge dual with respect to the spacetime
metric $\gtens$. The horizon quantities $M_i$  
and $Q_i$ are defined by the corresponding integrals over the 
topological $2$--spheres $\Hi$,
\be
M_i \, \equiv \, - \frac{1}{8 \pi} \int_{\Hi} \ast dk \; ,
\; \; \; \;
Q_i \, \equiv \, - \frac{1}{4 \pi} \int_{\Hi} \ast F \; .
\label{MiQi}
\ee
(It is worth noticing that the $M_i$ are not the same 
quantities as the $m_i$ appearing in the numerators of 
the PM solution. The latter are related to the masses of  
the black holes as seen from the asymptotically flat region, 
whereas the former can be expressed in terms of the area 
${\cal A}_i$ and the surface gravity $\kappa_i$ of $\Hi$; 
$M_i = \frac{1}{4 \pi} \kappa_i {\cal A}_i$.)

We shall now give an elementary derivation of the bound 
$M \geq |Q|$ in the general case, and of the equality 
$M =|Q|$ in the case of a degenerate  horizon. More 
precisely, we prove the following theorem: \\

\noindent
{\em Theorem:}
Let $\Mfold$ be an asymptotically flat electrovac spacetime 
with strictly static, simply connected domain of outer 
communications and not necessarily connected Killing horizon 
$H$. Let $M$ and $Q$ denote the total mass and charge, and 
let $V$ and $\phi$ be the gravitational and the electrostatic 
potential, respectively. Then
\begin{description}
\item{(i)}
$M \geq |Q|$;
\item{(ii)}
$M = |Q|$ if and only if 
$V = [\phi - \mbox{sig}(Q)]^2$, i.e., if and only if the metric 
is locally of the PM form;
\item{(iii)}
$M = |Q|$ if $H$ is completely degenerate and if all 
charges $Q_i$ (as defined in eq. (\ref{MiQi})) have the 
same sign;
\item{(iv)}
$M^2 = (\frac{1}{4 \pi} \kappa {\cal A})^2 + Q^2$ and
$V = \phi^2 - 2\frac{M}{Q} \phi + 1$ if $H$ is connected (with 
surface gravity $\kappa$ and area ${\cal A}$).
\end{description}

\noindent
{\em Proof:}
Let us start by compiling the relevant equations: In a static 
spacetime the vanishing of the twist implies that the electric 
and the magnetic fields, $E$ and $B$, are parallel with a constant 
factor of proportionality (see, e.g., \cite{CQG11-L49-94}).
One can therefore apply a duality rotation to achieve $B=0$.
(For our purposes it is convenient to use the $1$--forms 
$E \equiv -i_kF$ and $B \equiv i_k \ast F$, which should not be 
confused with the electric and the magnetic fields defined with 
respect to the normal of the spacelike hypersurface $\Sigma$.
In fact, the definitions coincide only if $\Sigma$ is a static
hypersurface, which will not be assumed in the following.)

In the static case the Maxwell equations reduce to
\be
d E \, = \, 0 \; , \; \; \; \; \cod \left( \frac{E}{V} \right) 
\, = \, 0 \; ,
\label{Maxwell}
\ee
where $\cod = \ast d \ast$ is the co--derivative with respect to 
the spacetime metric $\gtens$ and $i_k$ denotes  the interior
product with $k$. (For arbitrary $p$-forms $\Omega$, the latter
can be computed from the identity 
$i_k \Omega = - \ast (k \we \ast \Omega)$;
see, e.g., the appendix of \cite{CQG10-1299-93} or \cite{H-B} 
for Killing field identities). Since (under the natural 
requirements mentioned above \cite{CQG11-L147-94}) the domain 
of outer communications $\doc$ is simply connected, Maxwell's
equations imply the existence of a potential $\phi$, (with 
$E = d \phi$). Using the Einstein equation
$R(k,k) = 8 \pi T(k,k) = \sprod{E}{E}$ and the Killing field 
identity (see, e.g., \cite{H-B}) 
\be
\cod \left( \frac{dV}{V} \right) \, = \, - \, \frac{2}{V} R(k,k) \; ,
\label{Ric}
\ee
we obtain the (Poisson) equations for the potentials $\phi$ and $V$:
\be
\cod \left( \frac{d \phi}{V} \right) \, = \, 0 \; ,
\; \; \; \;
\cod \left( \frac{dV}{V} \right) \, = \, - \, \frac{2}{V} \, 
\sprod{d \phi}{d \phi} \; .
\label{Poisson}
\ee

The reasoning is now based on similar divergence identities as 
used by Israel in the first proof of the electrovac uniqueness 
theorem \cite{CMP8-245-68} (see also \cite{GRG5-61-72}). In order 
to derive these identities, it is convenient to apply  Stokes'
theorem in the form
\be
\int_{\partial \Sigma} f \, \ast (k \we \alpha) \, = \,
\int_{\Sigma} \left( \sprod{\alpha}{df} 
\, - \,  f \, \cod \alpha \right) \, \ast k \; ,
\label{Stokes}
\ee
where $f$ and $\alpha$ denote an arbitrary stationary function 
($L_k f = 0$) and an arbitrary stationary $1$-form 
($L_k \alpha = 0$), respectively. (The above formula is
obtained from integrating the dual of the identity
$\cod (k \we \Omega) = - (\cod \Omega) \, k$, which is valid for  
arbitrary stationary $1$-forms $\Omega$. One then chooses 
$\Omega = f \alpha$ and applies the product formula 
$\cod (f \alpha) = f \, \cod\alpha - \sprod{df}{\alpha}$.)
In order to be able to identify the boundary integrals with the 
quantities $M$, $Q$, $M_i$ and $Q_i$, we also note that in the 
purely electric, static case one has
\be
dk \, = \, - k \we \frac{dV}{V} \; ,
\; \; \; \;
F \, = \, k \we \frac{d \phi}{V} \; .
\label{dk-F}
\ee
(Staticity is equivalent to $k \we dk = 0$. Applying $i_k$ on this  
identity and using $i_k \, dk = -d i_k k = d V$ yields the above 
formula for $dk$.)

We are now in a position to derive the desired identities: First
choosing $\alpha = dV/V$ and then $\alpha = d\phi/V$ and using eqs.
(\ref{Poisson}) on the r.h.s. of Stokes' identity (\ref{Stokes}) and  
eqs. (\ref{dk-F}) on the l.h.s., we obtain the
identities ($\etatil \equiv i_k \eta = \ast k$)
\bea
8 \pi \, \left( \finf M - \sum f_i M_i \right) & = & 
\int_{\Sigma} \left( \frac{\sprod{dV \!}{df}}{V}
\, +  \, 2 \, f \, \frac{\sprod{E}{E}}{V} \right) \, \etatil \; ,
\label{identM} \\
4 \pi \, \left( \ginf Q - \sum g_i Q_i \right) & = & \, -
\int_{\Sigma} \frac{\sprod{E}{dg}}{V} \, \etatil \; ,
\label{identQ}
\eea
provided that $f$ and $g$ assume constant values on $\Sinf$ and on  
each component $\Hi$. This is always the case if $f$ and $g$ are 
algebraic expressions in terms of the potentials $\phi$ and $V$, 
since the latter are constant on $\Sinf$ and $\Hi$,
\bea
V \mid_{\Sinf} & = & 1 \; , \; \; \; \; V \mid_{\Hi} \, = \, 0 \; ,
\label{V-vals}\\
\phi \mid_{\Sinf} & = & 0 \; , \; \; \; \;
\phi \mid_{\Hi} \, = \, \phi_i \, = \, \mbox{const.}
\label{Phi-vals}
\eea
(We recall that this is a consequence of asymptotic flatness and the
general properties of Killing horizons. The latter also imply that  
$R(k,k) = 0$ on $H$, i.e., that $E$ is null on $H$. Since, by 
definition, $E$ is orthogonal to the null generator of the horizon, 
$E$ and $k$ are also parallel, $E = \sigma_i k$ on $\Hi$
(for some functions $\sigma_i$). Thus
$L_{\xi} \phi = i_{\xi} E = \sigma_i \sprod{k}{\xi} = 0$ on $\Hi$ 
for every tangential field $\xi$, implying that 
$\phi_i = \mbox{const.}$ Eventually, $\phi_{\infty} = 0$ follows from
asymptotic flatness and the gauge freedom.)

In order to prove (i), we choose $f[\phi, V] = \frac{1}{2}
\sqrt{V}$ and $g[\phi, V] = V$. Using eq. (\ref{V-vals}), the  
l.h.sides of the identities (\ref{identM}) and (\ref{identQ}) 
assume the values $4 \pi M$ and $4 \pi Q$, respectively.
We therefore obtain
\be
4 \pi \, \left( M  \pm Q \right) \, = \, \int_{\Sigma} 
\sprod{( E \mp \frac{dV}{2 \sqrt{V}})}{( E \mp \frac{dV}{2 \sqrt{V}})}
\; \frac{\etatil}{\sqrt{V}} \; .
\label{M-Q-ineq}
\ee
Since both $E$ and $dV$ are nowhere timelike in $\doc$, this already
establishes $M \geq |Q|$. (Here we need a strictly static domain of  
outer communications.) 

Clearly $M = |Q|$ holds if and only if the 
integrand in eq. (\ref{M-Q-ineq}) vanishes, that is, 
if $d \phi = \pm d \sqrt{V}$. The 
asymptotic values of the potentials then imply that 
$V = (\phi \pm 1)^2$ 
(and thus $\phi_i = \mp 1 \, , \forall i$). Considering now 
eq. (\ref{identQ}) for $g=1$ and for $g = \phi$, one finds 
$\mbox{sig}(\phi_i) = \mbox{sig}(Q)$.
As for the second statement in (ii), 
it was already 
observed by Majumdar \cite{PR72-390-47} that staticity, together with
the above relation between the potentials, implies that the metric is 
locally of the PM form. In order to see this, one first notes
that the Ricci tensor for the conformal spatial metric 
$\htens \equiv V^{-1} \gtiltens$ reads
\bdm
R^{(h)}_{ij} \, = \, R_{ij} \, - \, g_{ij} \, \frac{1}{2 V}
\left( \Delta V - \frac{\sprod{dV \!}{dV}}{V} \right) \, + \,  
\frac{1}{2 V^2} \, V,_{i} V,_{j} \, ,
\edm
where all quantities on the r.h.s. refer to the spacetime metric. Now  
using the consequence $V \Delta V = \frac{3}{2} \sprod{dV \!}{dV}$ of 
the field equations (\ref{Poisson}) and $V = (\phi \pm 1)^2$, as well
as the Einstein equations 
$R_{ij} = V^{-1} (g_{ij} \sprod{d \phi}{d \phi} - 2 \, \phi,_i \phi,_j)$,
enables one to conclude that $R^{(h)}_{ij}$ vanishes. Hence, the
spatial manifold with metric $V^{-1} \gtiltens$ is flat, implying the
second assertion in (ii).

In order to establish (iii), we add the identities (\ref{identM}) and
(\ref{identQ}) and determine the functions $f$ and $g$ such that the  
entire r.h.s. vanishes, i.e., such that
\be
2 f_{\infty}M \, - \, 2 \sum f_i M_i \, +
\, g_{\infty} Q \, - \, \sum g_i Q_i \, = \, 0 \; .
\label{M-Q-relation}
\ee
This is the case if $f$ and $g$ are subject to the differential  
equations
$\partial_{\phi} f = \partial_V g$, $2 f = \partial_{\phi} g$ and
$\partial_V f = 0$, the solutions of which are
\bea
f & = & a \, \phi \, + \, b \; ,
\nonumber\\
g & = & a \, (V + \phi^2) \, + \, 2 b \, \phi \, + \, c \; ,
\label{sol}
\eea
with arbitrary constants $a$, $b$ and $c$. Using 
the asymptotic values
$f_{\infty} = b$, $g_{\infty} = a + c$ 
and the horizon values 
$f_i = a \phi_i + b$, $g_i = a \phi_i^2 + 2b \phi_i + c$, 
eq. (\ref{M-Q-relation}) yields 
the following three relations (for $a=b=0$, $c=1$ and cyclic):
\bea
Q & = & \sum Q_i \; ,
\label{i1} \\
M & = & \sum \, \left( \, M_i \, + \, Q_i \phi_i \, \right) \; ,
\label{i2}\\
Q & = & \sum \, \left( \, 2 M_i \phi_i \, + 
\, Q_i \phi_i^2 \, \right) \; .
\label{i3}
\eea
Taking advantage of these identities, it is not hard to verify that
\be
M^2 \, = \, Q^2 \, + \, \sum \frac{Q}{Q_i} M_i^2 \, - \, \sum Q Q_i \, 
\left( \, \phi_i + \frac{M_i}{Q_i} - \frac{M}{Q} \, \right)^2 \; .
\label{mainrel}
\ee
Since, by definition, $M_i = \frac{1}{4 \pi} \kappa_i {\cal A}_i$,  
the second term on the r.h.s. vanishes if the horizon has only 
degenerate components; $\kappa_{i} = 0, \, \forall i$.
Also requiring that all charges $Q_i$ have the same sign, we find
\be
M^2 \, \leq Q^2 \; ,
\; \; \; \; \mbox{for} \; \; \kappa_i = 0 \; \;
\mbox{and} \; \; \mbox{sig}(Q_i) = \mbox{sig}(Q_j), \; \forall \, i,j \, .
\label{MlessQ}
\ee
In order to avoid a contradiction to (i), we conclude that $M = |Q|$.

Finally, with regard to (iv), we observe that eqs. 
(\ref{i1}) and (\ref{i2}) 
imply $Q = Q_H$ and $\phi_H = (M - M_H)/Q$, respectively, provided 
that the horizon has exactly one (non--degenerate or degenerate) 
component. Hence, the last term in eq. (\ref{mainrel}) does not 
contribute in this case, and the latter yields 
\be
M^2 \, = \,  Q^2 \, + \,  M_H^2 \, .
\label{MMHQ}
\ee 
In order to conclude the proof, we introduce the function ${\cal F}$, 
defined by
\be
{\cal F} \, = \, V \, - \, (\phi^2 - 2\frac{M}{Q} \phi + 1) \; .
\label{F}
\ee
Clearly ${\cal F} \mid_{\Sinf} = 0$. By virtue of eq. (\ref{MMHQ})
and the above expression for $\phi_H$ we also have ${\cal F}_H = 0$, 
since ${\cal F} = V - (\phi - \phi_{H})(\phi - \phi_{H}^{-1})$.
In addition, the field equations (\ref{Poisson}) for the potentials  
imply $\cod(\frac{d{\cal F}}{V}) = 0$. Hence, Stokes' theorem  
(\ref{Stokes}) (with $f={\cal F}$ and $\alpha = d{\cal F}/V$) yields
\be
\int_{\Sigma} \frac{\sprod{d{\cal F}}{d{\cal F}}}{V} \; 
\etatil \, = \, 0 \; ,
\label{dFdF}
\ee
from which we conclude that ${\cal F}$ is constant. Thus, ${\cal F}$  
vanishes identically, i.e., $V = \phi^2 - 2\frac{M}{Q} \phi + 1$,
which concludes the proof of the theorem. \\

\noindent
{\em Concluding comments:}
The inequality (i), $M \geq |Q|$, has been established by Gibbons 
{\it et al} under more general conditions \cite{CMP88-295-83} (see 
also \cite{PLB109-190-82}). The general derivation proceeds along 
similar lines as Witten's proof of the positive mass theorem 
\cite{CMP80-381-81}. In particular, the proof requires solutions of 
the Dirac (Witten) equation with appropriate fall--off and
boundary conditions \cite{PLB121-241-83}. Since not all problems  
concerning the existence of such spinor fields have been settled in 
a completely conclusive manner, we feel that the elementary
derivation of $M \geq |Q|$ given above -- although restricted to the
static multi black hole case -- might be of certain use.

It is worth recalling that the property (iv) -- which was 
originally derived for a non--degenerate horizon -- constitutes the  
backbone of the uniqueness proof for the Reis\-sner--Nord\-str\"om  
solution. The fact that $\phi$ depends only on the gravitational 
potential $V$ was first established by Israel \cite{CMP8-245-68} and 
later (under weaker assumptions) by M\"uller zum Hagen
{\it et al} \cite{GRG5-61-72}. We also note that in the
case of a connected horizon, 
(i) and (ii) are immediate consequences of (iv).

Regarding the main result (iii), I am not aware of an 
alternative proof establishing $M = |Q|$ for the multi black hole 
case. The question whether this  
relation also holds if at least one (instead of each) component of 
$H$ is degenerate is presently under investigation. In addition, it 
would be desirable to find a proof which works for arbitrary signs of
the horizon charges $Q_i$. Eventually, a generalization of the above 
results to the stationary (not static) realm is also needed. \\

\noindent
{\em Acknowledgments:}
I wish to thank Piotr Chru\'sciel for numerous suggestions
and valuable comments on a first draft of this work.
I also acknowledge discussions with the Relativity Groups at the
University of Zurich and the Enrico Fermi Institute.
This work was supported by the Swiss National Science Foundation.

\end{document}